\documentclass[twocolumn,times]{aastex62}

\usepackage{textcomp}
\usepackage{amssymb}
\usepackage{graphicx,graphics}			
\usepackage{epsfig}						
\usepackage{dcolumn}					
\usepackage{bm}							
\usepackage{natbib}
\usepackage{times}
\usepackage{ulem}
\newcommand{\hi}{H{\sc i}}
\newcommand{\hii}{H{\sc i} 21\,cm}
\newcommand{\kms}{km~s$^{-1}$}
\newcommand{\rhom}{\rho_{HI}/\rho_{c,0}}

\shorttitle{H{\sc i} in star-forming galaxies at $0.2<z<0.4$}
\shortauthors{Bera et al.}
\begin{document}
\title{Atomic hydrogen in star-forming galaxies at intermediate redshifts}

\correspondingauthor{Apurba Bera}
\email{apurba@ncra.tifr.res.in}

\author{Apurba Bera}
\affiliation{National Centre for Radio Astrophysics, Tata Institute of Fundamental Research, Ganeshkhind, Pune - 411007, India}

\author[0000-0002-9757-7206]{Nissim Kanekar}
\altaffiliation{Swarnajayanti Fellow}
\affiliation{National Centre for Radio Astrophysics, Tata Institute of Fundamental Research, Ganeshkhind, Pune - 411007, India}

\author{Jayaram N. Chengalur}
\affiliation{National Centre for Radio Astrophysics, Tata Institute of Fundamental Research, Ganeshkhind, Pune - 411007, India}

\author{Jasjeet S. Bagla}
\affiliation{Indian Institute of Science Education and Research Mohali, Knowledge City, Sector 81, Sahibzada Ajit Singh Nagar, Punjab 140306, India}

\begin{abstract}
We have used the upgraded Giant Metrewave Radio Telescope to carry out a deep (117 on-source hours) 
L-band observation of the Extended Groth Strip, to measure the average neutral hydrogen 
(H{\sc i}) mass and median star formation rate (SFR) of star-forming galaxies, as well 
as the cosmic H{\sc i} mass density, at $0.2 < z < 0.4$. This was done by stacking the H{\sc i} 
21\,cm emission and the rest-frame 1.4~GHz radio continuum from 445 blue star-forming galaxies 
with $\rm M_B \leq -17$ at $z_{\rm mean} \approx 0.34$. The stacked H{\sc i} 21\,cm emission signal 
is detected at $\approx 7\sigma$ significance, implying an average H{\sc i} mass of 
$\rm \langle M_{HI} \rangle = (4.93 \pm 0.70) \times 10^9 \: M_{\odot}$. We also stacked the rest-frame 
1.4~GHz radio continuum emission of the same galaxies, to obtain a median SFR of 
$(0.54 \pm 0.06) \: {\rm M}_\odot$~yr$^{-1}$; this implies an average atomic gas depletion time 
scale of $\rm \langle \Delta t_{HI}\rangle \approx$ 9~Gyr, consistent with values in star-forming 
galaxies in the local Universe. This indicates that the star-formation efficiency does not 
change significantly over the redshift range $0 - 0.4$. We used the detection of the 
stacked H{\sc i} 21\,cm emission signal to infer the normalized cosmic H{\sc i} mass density 
$(\rm \rho_{HI}/\rho_{c,0})$ in star-forming galaxies at $z \approx 0.34$. Assuming 
the local relation between H{\sc i} mass and absolute B-magnitude, we obtain 
$\rm \rho_{HI}/\rho_{c,0} = (4.81 \pm 0.75) \times 10^{-4}$, implying no significant
evolution in $\rm \rho_{HI}/\rho_{c,0}$ from $z \approx 0.4$ to the present epoch.  
\end{abstract}



\section{Introduction}

Understanding galaxy evolution requires us to understand evolution in the two main baryonic constituents 
of galaxies, the stars and the gas. Over the last two decades, much progress has been made in understanding 
the evolution of the stellar properties of galaxies 
\citep[e.g. the star formation rate (SFR), the main sequence, etc; ][]{hopkins06apj,noeske07apjl}, 
via optical and near-infrared studies of the so-called ``deep fields''. Our understanding  
of molecular gas in high-$z$ galaxies has also improved, via blind and targetted CO studies 
\citep[e.g.][]{tacconi18apj,pavesi18apj,decarli19apj}. In contrast, little is known about 
{\it atomic gas}, 
primarily neutral hydrogen (\hi), in high-$z$ galaxies, although this is the primary fuel reservoir for star 
formation.
Understanding the redshift evolution of the \hi\ content of star-forming galaxies is essential to 
obtain a complete picture of galaxy evolution.

\begin{figure*}[t!]
\centering
\includegraphics[scale=0.5,trim={0.0cm 1.0cm 1cm 3.5cm},clip]{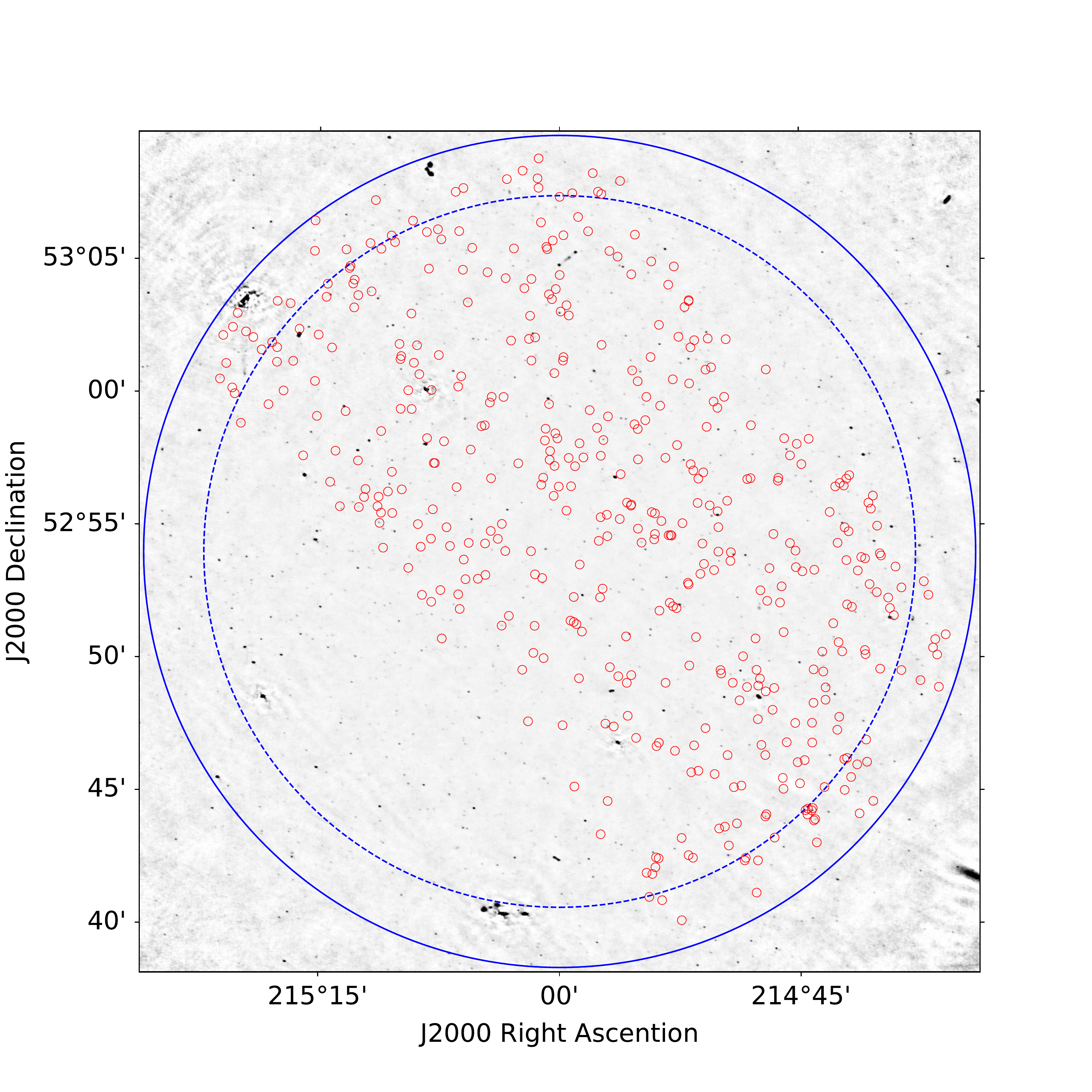}
\caption{The uGMRT 1.2~GHz image of the EGS. The solid and dashed blue circles indicate 
the FWHM of the uGMRT primary beam at 1014~MHz ($z = 0.4$) and 1183~MHz ($z = 0.2$), respectively. The 
red circles indicate the locations of the 445 galaxies of our sample.}
\label{fig:continuum}
\end{figure*}

In the local Universe, \hii\ emission surveys have yielded the \hi\ masses of thousands of galaxies 
\citep[e.g.][]{zwaan05mnras,jones18mnras}, providing accurate measurements of the \hi\ mass function 
and the cosmic \hi\ mass density. Unfortunately, it is difficult 
to detect the weak \hii\ line from galaxies at $z \gtrsim 0.2$ with current telescopes 
\citep[e.g.][]{jaffe13mnras,catinella15mnras}, with 
only one galaxy detected at $z \gtrsim 0.25$, at $z \approx 0.376$ in the Very Large 
Array CHILES survey of the COSMOS field \citep[][]{fernandez16apjl}.

While detecting individual galaxies in \hii\ emission at $z \gtrsim 0.3$ will require very deep 
integrations in the foreseeable future, 
one can measure the average \hi\ content of samples of galaxies within the primary beam of a radio 
interferometer, by ``stacking'' their \hii\ emission signals \citep[e.g.][]{chengalur01aa}. 
Of course, such stacking requires large galaxy samples with accurate redshifts, usually from optical 
spectroscopy. This requirement has meant that, despite multiple studies, there are still no 
statistically-significant detections of the stacked \hii\ emission signal at $z \gtrsim 0.2$ 
\citep[e.g.][]{lah07mnras,kanekar16apjl,rhee18mnras}.

The paucity of information on the \hi\ content of galaxies at even relatively low redshifts, $z \gtrsim 0.25$, 
has motivated us to begin an \hii\ ``deep field'' with the upgraded Giant Metrewave Radio Telescope (uGMRT).
We chose the Extended Groth Strip (EGS) as the target field, due to the outstanding ancillary data in the 
EGS over a wide range of wavelengths, as well as the excellent spectroscopic coverage in this field 
from the DEEP2 and DEEP3 surveys \citep{newman13apjs,cooper12mnras}. In this {\it Letter}, we present 
results from the first part of this uGMRT \hii\ emission survey\footnote{We use a Lambda Cold Dark Matter ``737'' cosmology, with $\Omega_m = 0.3$, $\Omega_\Lambda=0.7$, and $H_0$=70 km s$^{-1}$  Mpc$^{-1}$, throughout this manuscript}.

\begin{figure*}[t!]
\centering
\includegraphics[scale=0.8,trim={0.5cm 0.5cm 0 0.25cm},clip]{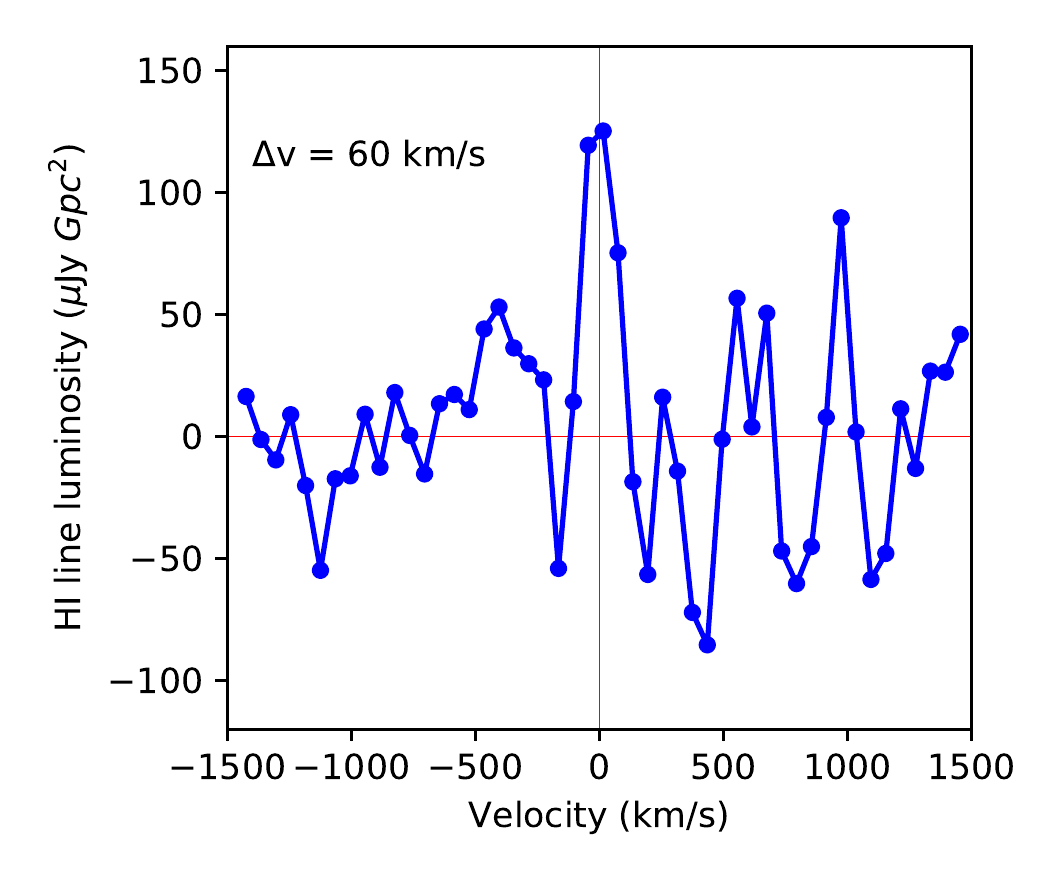}
\includegraphics[scale=0.8,trim={0.0cm 0.5cm 0 0.25cm},clip]{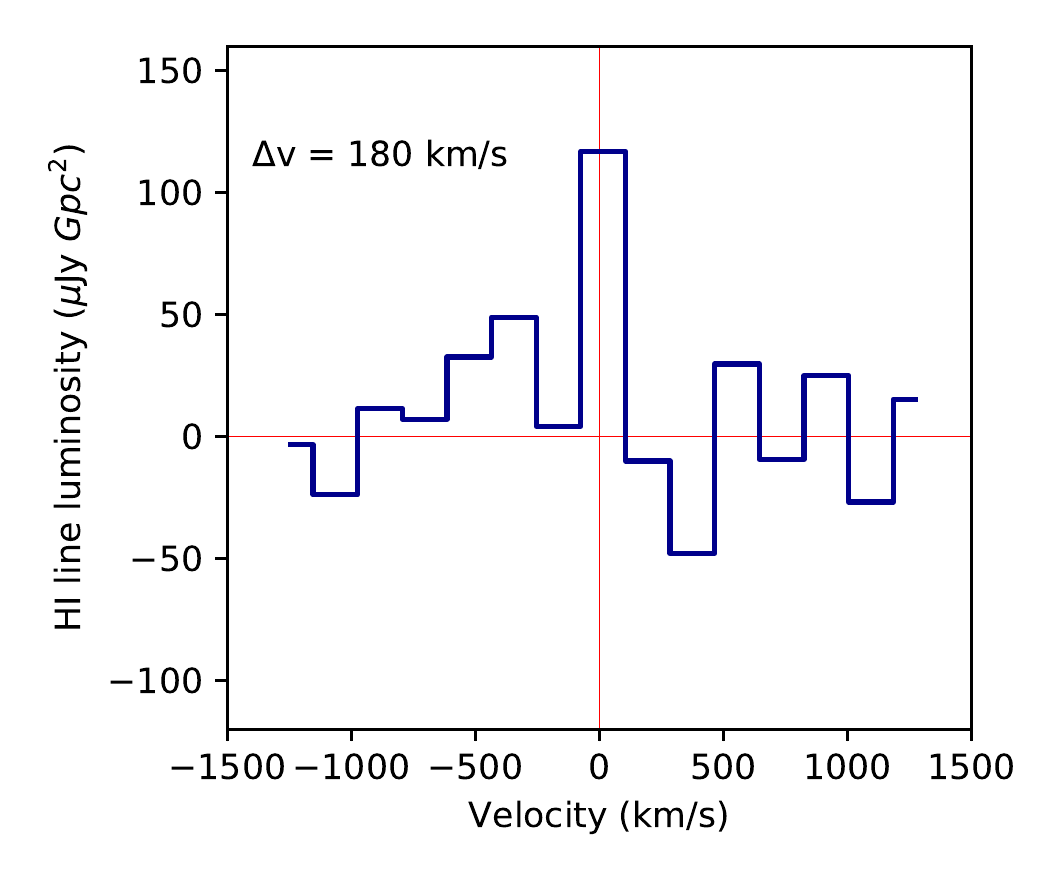}
\caption{The stacked \hii\ emission spectrum of the 445 galaxies of our sample, at 
velocity resolutions of $60$~\kms\ (left panel) and $180$~\kms (right panel).}
\label{fig:spectrum}
\end{figure*}

\section{Observations \& Data Analysis}

We used the uGMRT L-band receivers to carry out a deep integration on the EGS between March 2017 
and April~2018 (proposals 31\_038 and 34\_083). The total integration time was $\approx 175$~hours, 
with an on-source time of $\approx 117$~hours. The pointing centre was chosen to be RA=14h20m00.0s, 
Decn.=+52$^{\circ}$54'00.0", close to the centre of the region with the richest multi-wavelength 
coverage. The GMRT Wideband Backend was used as the correlator, with a bandwidth of 
400~MHz, sub-divided into 8192 channels, and two polarizations. The 400~MHz band covers the frequency 
range $970-1370$~MHz, corresponding to $z \approx 0.037-0.46$, with a velocity resolution of 
$\approx 10-15$~\kms\ across the band.

The initial analysis, including data editing, and gain and bandpass calibration was carried 
out using ``classic'' {\sc aips}, following standard procedures. After the initial calibration, 
the 400~MHz dataset was split into four 100-MHz sub-bands, and a standard self-calibration procedure 
\citep[e.g.][]{kanekar16apjl} was then run independently in {\sc aips} on each sub-band. 
The self-calibrated data of the four sub-bands were then stitched together, and the combined visibilities 
imaged in {\sc casa}, using {\sc tclean}.
An area of $\approx 1^{\circ} \times 1^{\circ}$ was imaged, out to the first null of the uGMRT primary beam. 
A single final amplitude-and-phase self-calibration was then carried out to bring the 
different sub-bands onto the same flux density scale, and a final continuum image then made from the 
self-calibrated data, using {\sc tclean}, with Briggs weighting (robust=0.5). This image, shown in Fig.~\ref{fig:continuum}, 
has a resolution of $\approx 3.5'' \times 2.7''$ and a root-mean-square (RMS) 
noise of $\rm \approx 2.3\: \mu Jy$/Bm near the image centre. 

The gain solutions obtained from the self-calibration procedure were then applied to the spectral 
visibilities, and the continuum image subtracted out, before a final round of RFI excision 
was carried out on the residual data. The residual multi-channel visibilities were then shifted 
to the barycentric frame and imaged in {\sc casa}, using natural weighting.
This spectral cube was corrected for the (frequency-dependent) shape of the uGMRT primary beam. 
Small sub-cubes were extracted at the location of each target galaxy, shifted to velocity,
in the rest frame of the galaxy, and smoothed to, and re-sampled at, a velocity resolution of 60~\kms.



\section{Stacking the \hii\ emission and the radio continuum}

To obtain a clean interpretation of the \hii\ stacking results, it is important to use a 
uniformly-selected sample of galaxies, with accurate redshifts (redshift errors~$\lesssim 100$~km~s$^{-1}$). 
For the EGS, the DEEP2 and DEEP3 surveys have a redshift accuracy of $\approx 30$~\kms\ 
\citep{newman13apjs}, sufficient for \hii\ stacking. We restricted our target sample to ``blue''
star-forming galaxies, using the colour division of \citet{coil08apj} between red and blue galaxies,
[$\rm (U-B)=-0.032 \times (M_B + 21.62) + 1.035$]. We also used a uniform absolute B-magnitude 
limit to select our targets, with $M_B \leq -17$; with the DEEP2 and DEEP3 spectroscopic 
criteria, this yields a near-complete, absolute-magnitude-limited sample of galaxies at $0.2 < z < 0.4$. 
Finally, the galaxies were selected to lie within the full-width-at-half-maximum (FWHM) of the uGMRT primary 
beam at the frequency of the redshifted \hii\ line, to have redshift quality code $>3$ 
\citep[i.e. reliable redshifts; ][]{newman13apjs}, and to not be classified as active galactic 
nuclei in either the DEEP2/DEEP3 catalogs \citep[based on the best-fit template to the optical spectral energy distribution][]{cooper12mnras,newman13apjs} or detected in our radio continuum image, with 
a 1.4~GHz luminosity $\geq 2 \times 10^{23}$~W~Hz$^{-1}$ \citep[e.g.][]{condon02aj}.

We do not {\it a priori} know the optimal spatial resolution at which to extract the spectrum 
for each galaxy. Using too high a resolution would resolve out some of the \hii\ emission, 
thus reducing the signal-to-noise ratio (S/N). Conversely, using too coarse a resolution would 
imply that only some of the interferometric baselines (the shorter ones) are being used to extract 
the spectrum, again lowering the S/N. We hence extracted spectra at a range of spatial resolutions, 
$20-100$~kpc at the redshift of each galaxy, 
and stacked the spectra at each resolution, to determine the optimal spatial resolution. At 
each resolution, an error spectrum was also extracted for each galaxy, by measuring the RMS noise 
per channel in the sub-cube of the galaxy, from an annular region that excludes the galaxy itself. 
A second-order baseline was then fitted to each spectrum, using the velocity range $\pm 1500$~\kms\ 
around the galaxy redshift, and subtracted out.

Before stacking, we used the Kolmogorov-Smirnov and Anderson-Darling tests to test the spectra for Gaussianity, 
excluding spectra with p-values $< 0.002$ in either test. Our final sample contains 445 blue star-forming 
galaxies with $\rm M_B \leq -17$, at $0.2 < z < 0.4$, and with an average redshift of $\langle {z}\rangle = 0.34$.


\begin{figure*}
\centering
\includegraphics[width=3.5in]{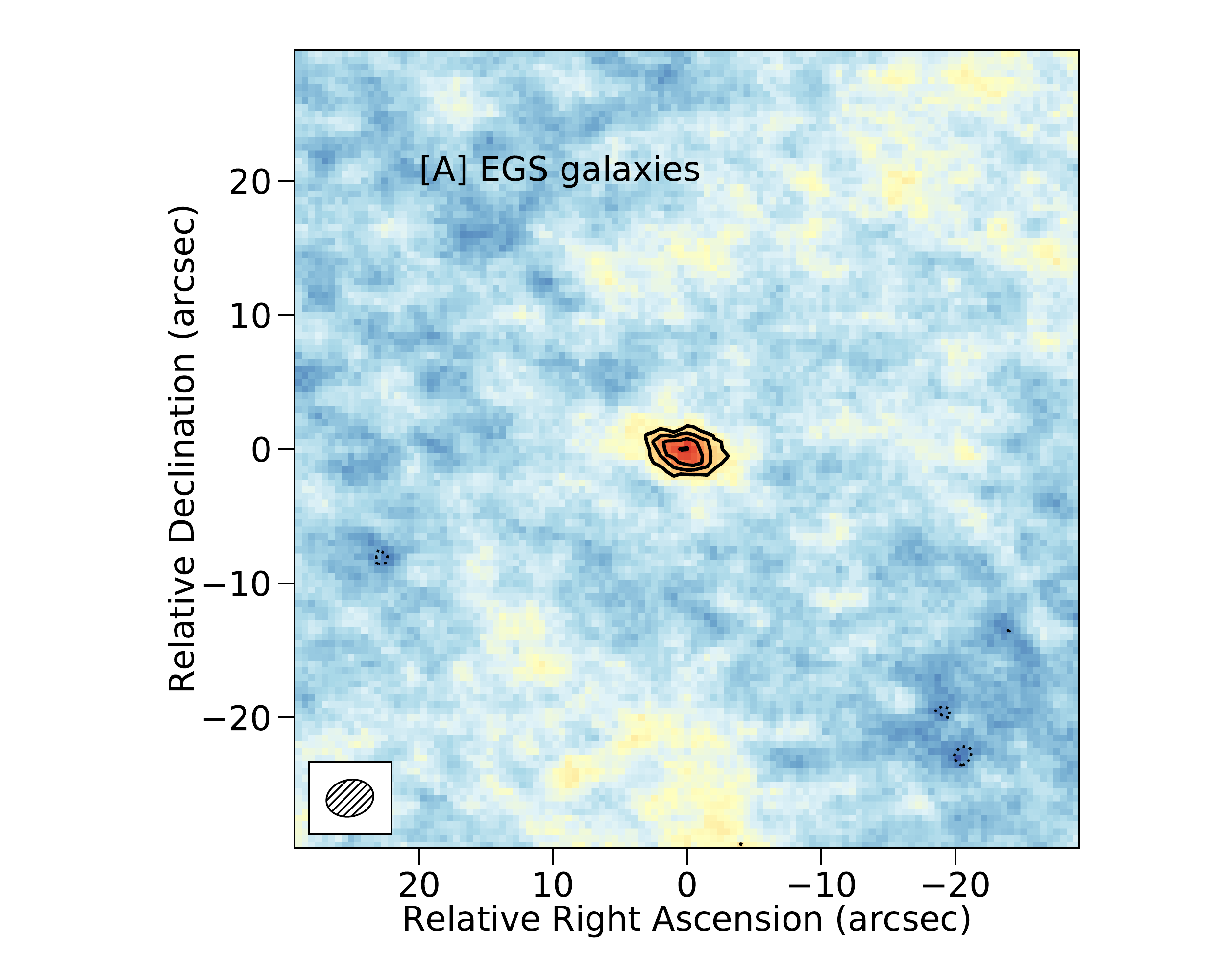}
\includegraphics[width=3.5in]{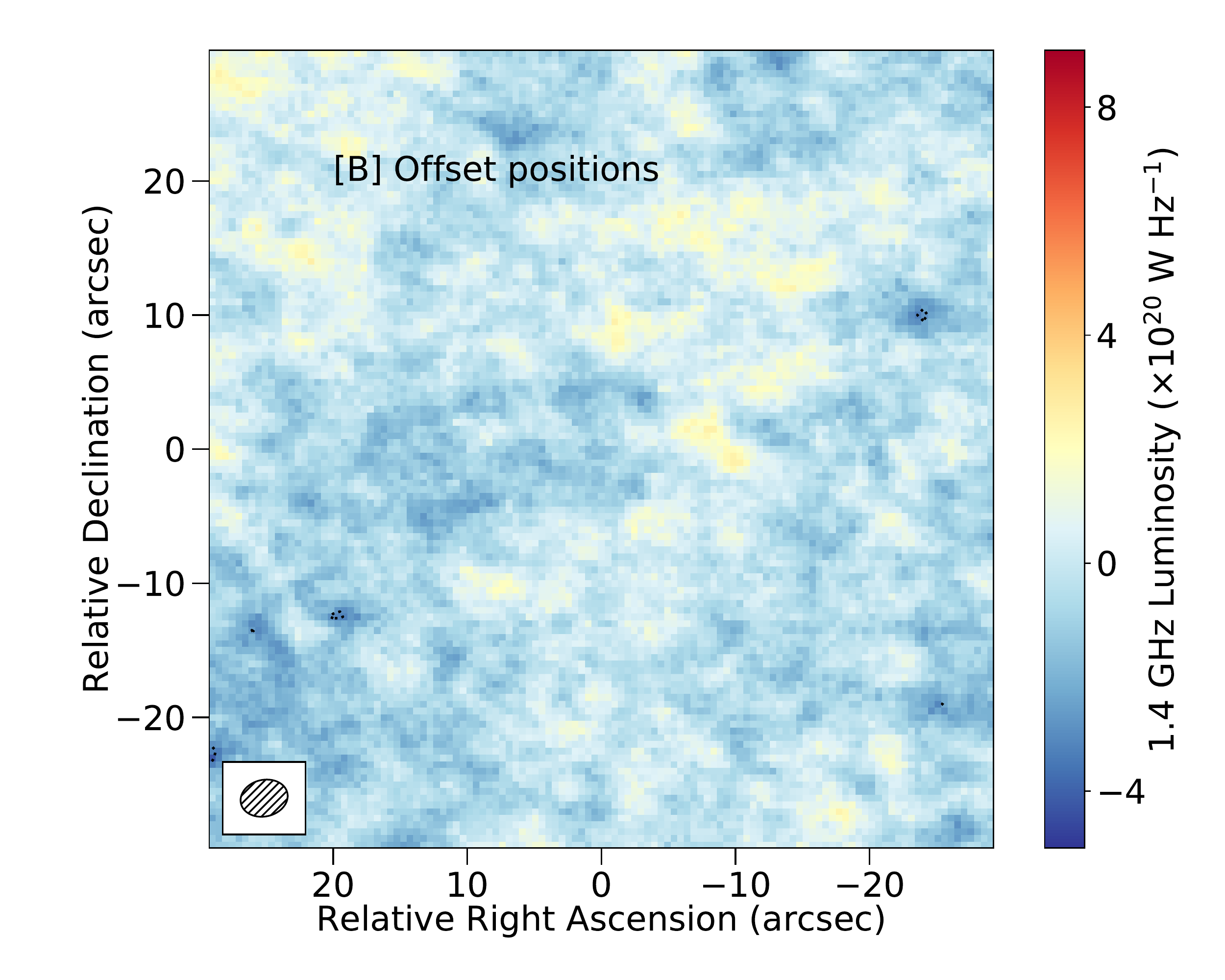}
\caption{The stacked 1.4~GHz luminosity image (in units of $\rm 10^{20}\:W\:Hz^{-1}$) 
from [A]~(left panel) the 445 galaxies of our sample, and [B]~(right panel) offset locations
$50^{''}$ away from the sample galaxies. The synthesized beam is shown as an ellipse in 
the bottom left.
}
\label{fig:continuum-stack}
\end{figure*}

The stacking was carried out in units of \hii\ line luminosity (rather than flux density), 
to account for the spread in the luminosity distances of the target galaxies. While the uGMRT sensitivity 
is approximately constant (within $\approx 10$\%) across the observing band, the RMS noise on the 
flux density spectra varies due to both the galaxy locations in the primary beam and 
frequency-dependent data flagging. In addition, the RMS noise in the luminosity density spectra 
increases with increasing redshift. 
The optimal approach of weighting by the inverse of the variance of the luminosity density spectra 
would imply significantly lower weights for the higher-redshift galaxies, i.e. a bias towards lower redshifts. 
We hence chose to weight each spectrum by the inverse of its variance in {\it flux density}, so that the 
effective redshift of the stacked spectrum remains the average redshift of the sample, $\langle {z} \rangle =0.34$.



Our final stacked \hii\ spectrum, at a resolution of 40~kpc, is shown in Fig.~\ref{fig:spectrum}. The spectrum 
shows a clear detection of \hii\ emission, with an integrated luminosity density of 
$\rm L_{HI} = (2.10 \pm 0.30) \times 10^4 \: Jy\: Mpc^2 \: km \: s^{-1}$. The signal was also detected in 
the stacks at resolutions of 20~kpc and 30~kpc, but with a lower integrated luminosity density, indicating 
that the total  \hii\ emission has not been recovered. Using a spatial resolution coarser than 40~kpc does 
not significantly change the integrated luminosity density, but lowers the S/N due to an increased RMS noise.

We also stacked the radio continuum images of our 445 galaxies, to measure their average SFR 
from the 1.4~GHz radio luminosity \citep[e.g.][]{yun01apj}. We followed the procedure of \citet{bera18apj}, 
using ``median stacking'' to stack the 1.4~GHz radio luminosity images of the sample.
A radio spectral index of $\alpha = -0.8$ (with flux density $S_\nu \propto \nu^\alpha$) was assumed, 
to shift each galaxy's image from the observing frequency of 1.2~GHz to a rest-frame frequency of 1.4~GHz.

Figure~\ref{fig:continuum-stack}[A] shows the stacked 1.4~GHz continuum image of the 445 galaxies.
A marginally resolved source is clearly detected, with a 1.4~GHz luminosity of $\rm L_{1.4\;GHz} = 
(1.45 \pm 0.18) \times 10^{21}\:W \: Hz^{-1}$. No emission was seen in a stack of positions offset by 
$50''$ from the sample galaxies, to test for possible systematic effects (see figure 3[B]; e.g. \citealt{bera18apj}).


\section{Results \& discussion}

\subsection{The atomic gas mass and gas fraction}


For optically-thin \hii\ emission, the \hi\ mass of a galaxy is related to its velocity-integrated 
\hii\ line luminosity density by the relation 
\begin{equation}
\rm M_{HI} = 2.343 \times 10^5 \times \int L_{HI} dv \; ,
\end{equation}
where $\rm M_{HI}$ is in units of $\rm M_\odot$ and $\rm L_{HI}$ in $\rm Jy\:Mpc^2\:km\:s^{-1}$. Our measured 
stacked \hii\ line luminosity density of $\rm L_{HI} = (2.10 \pm 0.30) \times 10^4 \: Jy\: Mpc^2 \: km \: s^{-1}$ 
then yields an average \hi\ mass of $\rm \langle M_{HI} \rangle = (4.93 \pm 0.70) \times 10^9 \: M_{\odot}$, for the galaxies 
of the sample. The line FWHM is $\approx 200$~\kms.

In the local Universe, the \hi\ mass of a galaxy is correlated with its \hi\ diameter, $\rm D_{HI}$
\citep[e.g.][]{wang16mnras}. If the local $\rm D_{HI} - M_{HI}$ relation applies to galaxies at $z \approx 0.34$,
our average \hi\ mass of $\rm \approx 5 \times 10^9 \: M_\odot$ implies an average diameter of $\approx 40$~kpc.
This is consistent with our finding that a spatial resolution of 40~kpc recovers all the \hii\ emission.

To estimate the effect of source confusion on the measured average \hi\ mass, we identified galaxies in our 
sample with ``close companion(s)'', defined as a neighbour separated by $\leq 40$~kpc and $\leq 300$~\kms\ 
in the DEEP2/DEEP3 catalogues. Only $\lesssim 10\%$ of the galaxies of our sample have such ``close companion(s)'';
excluding these from the stack does not significantly change the average \hi\ mass. Source confusion thus has 
little effect on our measurement.

The average stellar mass of the galaxies of our sample is $\rm \langle M_* \rangle = 4.1 \times 10^9 \: M_\odot$ \citep{mostek12apj,stefanon17apjs}, 
implying a ratio of average \hi\ mass to average stellar mass, $\rm \langle M_{HI}\rangle/\langle M_*\rangle, \approx 1.2$. 
This is somewhat larger than values of $\rm M_{HI}/M_*$ in blue star-forming galaxies in the local Universe: 
for example, applying our stellar mass function to blue galaxies in the xGASS sample (with 
NUV-r~$\leq 4$; \citealp{catinella18mnras}) yields $\rm M_{HI}/M_* \approx 0.5$ (although we note that  
the xGASS sample has $M_* \geq 10^9 M_\odot$, somewhat larger than the stellar mass of our sample).


\subsection{The atomic gas depletion timescale}

The atomic gas depletion timescale, $\rm \delta t_{HI} = M_{HI}/SFR$, is the time 
taken by a galaxy to consume its \hi\ reservoir at its current SFR, and is a measure of 
its star-formation efficiency. A short \hi\ depletion time scale suggests that star formation is likely to 
soon be quenched, in the absence of gas inflow. In the local Universe, studies of stellar-mass-selected 
samples of star-forming galaxies (with, typically, 
$\rm M_* \gtrsim 10^9 \: M_\odot$) have obtained $\rm \Delta t_{HI} \approx 3-10$~Gyr, with little 
evidence for dependence on galaxy properties like stellar mass, galaxy colour, etc 
\citep[e.g.][]{schiminovich10mnras,wong16mnras,saintonge17apjs,catinella18mnras}. However, stacking the \hii\ emission and the radio continuum of star-forming galaxies at $z \approx 1.3$ (with $\rm M_B \lesssim -21$) 
yielded a far lower average \hi\ depletion time, $\rm \langle \Delta t_{HI} \rangle < 0.87$~Gyr 
\citep{kanekar16apjl,bera18apj}.  This suggests evolution in the \hi\ depletion time for 
star-forming galaxies from $z \approx 1.3$ to $z \approx 0$.

Our stack of the 1.4~GHz radio continuum emission of 445 blue star-forming galaxies yielded a median 1.4~GHz 
radio luminosity of $\rm L_{1.4\;GHz} = (1.45 \pm 0.18) \times 10^{21}\:W \: Hz^{-1}$. Using the relation 
$\rm SFR = 3.7 \times 10^{-22} \: W \:\; Hz^{-1}$ (assuming a Chabrier initial mass function; \citealt{yun01apj}) gives a median SFR of $\rm (0.54 \pm 0.06) \: M_{\odot}\; yr^{-1}$. Combining this with the average \hi\ mass 
yields $\rm \langle {\Delta t}_{HI}\rangle \approx 9$~Gyr, consistent with the values of $\rm \Delta t_{HI}$ 
in the local Universe, but far lower than that at $z\approx 1.3$. This suggests either 
evolution in the star-formation efficiency of galaxies between $z \approx 1.3$ and $z \approx 0.34$, 
soon after the epoch of galaxy assembly, or a much higher efficiency in brighter galaxies at $z \approx 1.3$.

\subsection{The cosmic \hi\ density}

\begin{figure}
\centering
\includegraphics[scale=0.72,trim={0.5cm 0.25cm 0 0},clip]{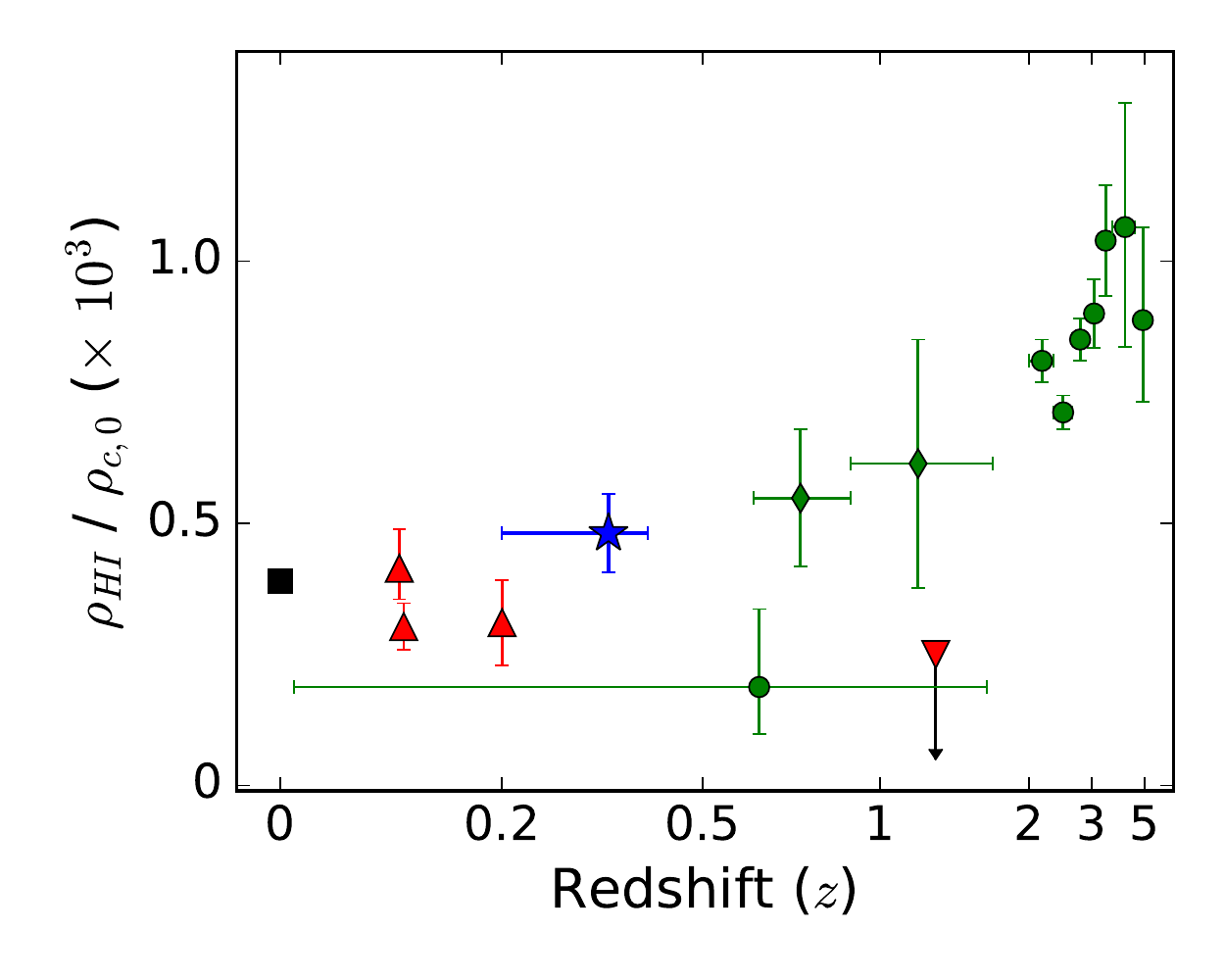}
\caption{The evolution of $\rm \rhom$ with redshift, with all values in a 737 cosmology \citep{neeleman16apj}. The black square is from blind \hii\ emission surveys in the local Universe \citep{jones18mnras}, the red triangles from earlier low-$z$ \hii\ stacking experiments \citep[with $\geq 3\sigma$ significance][]{rhee13mnras,delhaize13mnras}, the green circles from DLA studies \citep{noterdaeme12aa,crighton15mnras,neeleman16apj}, and the green diamonds 
from Mg{\sc ii} absorption samples \citep{rao17mnras}. The inverted red triangle shows the $3\sigma$ upper limit 
on $\rm \rhom$ in star-forming galaxies with $\rm M_B\lesssim-21$ at $z\approx1.3$ \citep{kanekar16apjl}. The blue star shows the present measurement, at $z \approx 0.34$.}
\label{fig:omegahi}
\end{figure}

Our estimate of the average \hi\ mass, $\rm \langle {M}_{HI} \rangle = (4.93 \pm 0.70) \times 10^9   \: M_\odot$,
of an absolute-magnitude-limited sample of galaxies can be used to infer the cosmic mass density of \hi\ in 
galaxies, $\rm \rho_{HI}/\rho_{c,0}$, at $z \approx 0.34$, where $\rm \rho_{c,0}$ is the 
comoving critical density at $z = 0$. To estimate $\rm \rhom$ in galaxies with $\rm M_B \leq -17$, 
we simply integrate over the B-band luminosity function at $z \approx 0.3$ for $\rm M_B \leq -17$, 
and multiply by $\rm \langle M_{HI} \rangle$. The B-band luminosity function at $z \approx 0.3$ is \citep{willmer06apj}
\begin{equation}
\rm \phi(L_B) = \left(\frac{\phi_*}{L_{B*}}\right)\left(\frac{L_B}{L_{B*}}\right)^{\alpha}\;e^{-L_B/L_{B*}}  \:,
\label{eqn-blum}
\end{equation} 
where $\rm \phi_*=(32 \pm 2) \times 10^{-4} Mpc^{-3}$, $\rm L_{B*}=(1.77 \pm 0.09) \times 10^9 \: L_{B\odot}$ 
and $\rm \alpha=-1.30$. Integrating over this luminosity function for $\rm M_B \leq -17$ yields a 
comoving galaxy number density of $0.013$~Mpc$^{-3}$. This yields $\rm \rho_{HI} = (6.41 \pm 0.91) 
\times 10^7 \: M_\odot \: Mpc^{-3}$, and $\rm \rhom = (4.75 \pm 0.67) 
\times 10^{-4}$ for galaxies with $\rm M_B \leq -17$ at $0.2 < z < 0.4$. We emphasize that this 
estimate of $\rm \rhom$ does {\it not} include contributions from galaxies fainter than $\rm M_B=-17$.

The total $\rm \rhom$ in blue galaxies at $z \approx 0.3$ may be estimated by combining the 
B-band luminosity function at $z \approx 0.3$ with the relation between $\rm M_{HI}$ and $\rm M_B$ 
\citep[e.g.][]{denes14mnras},
\begin{equation}
\rm \frac{M_{HI}}{L_B} = K \left(\frac{L_B}{L_{B*}}\right)^{\beta} \:,
\label{eqn-mhmb}
\end{equation}
where $\rm K$ is a constant equal to $\rm M_{HI}/L_B$ for $\rm L_B=L_{B*}$. Combining the above 
equations, and integrating over the B-band luminosity function (this time, for all $\rm M_B$), 
we obtain 
\begin{equation}
\rm \rhom = \frac{K}{\rho_{c,0}} \:L_{B*}\:\phi_*\:\Gamma(\alpha + \beta + 2, L_{min}/L_{B*}) \:,
\label{eqn-rhoh}
\end{equation}
where $\rm L_{min}$ is the faint-end cut-off of the B-band luminosity function. The result is 
insensitive to the choice of $\rm L_{min}/L_{B*}$ as long as this value is $\lesssim 10^{-3}$.  

There is currently no direct estimate of $\rm \beta$ at $z \approx 0.3$. We hence assume
that the local $\rm M_{HI}-M_B$ relation \citep[$\rm \beta = -0.15$; ][]{denes14mnras} is 
valid at $z \approx 0.3$. We then obtain $\rm K= (0.73 \pm 0.11) \: M_{\odot}/L_{B\odot}$.
We also include the effect of cosmic variance by using the estimated cosmic variance of 18\% in 
the luminosity density $\rm \phi_*$ at $z\approx 0.3$ for the EGS \citep{willmer06apj}.
Replacing for $\rm K$ in equation~(\ref{eqn-rhoh}), with $\rm L_{min}/L_{B*}=10^{-4}$, we obtain 
$\rm \rho_{HI} = (6.5 \pm 1.0 \pm 1.2)\times 10^7\:M_{\odot}\: Mpc^{-3}$ and 
$\rm \rhom =(4.81 \pm 0.75 \pm 0.87)\times 10^{-4}$, where the second uncertainty in the two 
expressions stems from cosmic variance. Note that using the $\rm M_{HI} - M_B$ relation 
to estimate $\rhom$ in galaxies with $\rm M_B \leq -17$ yields values of $\rhom$ consistent 
(within the statistical errors) with our estimate above.


In the local Universe, $\rm \rhom$ has long been accurately estimated via blind 
\hii\ emission surveys \citep[e.g.][]{zwaan05mnras,jones18mnras}. Conversely, at high redshifts, 
$z > 2$, the incidence of damped Lyman-$\alpha$ absorbers (DLAs) in QSO spectra from the Sloan Digital 
Sky Survey has been used to infer $\rm \rhom$ \citep[e.g.][]{prochaska05apj,noterdaeme12aa}. 
These studies have shown that $\rm \rhom$ declines by only a factor of 
$\approx 2$ from $z \approx 2.2$ to $z \approx 0$. However, the need for ultraviolet spectroscopy 
to detect DLAs at $z < 1.7$ and the weakness of the \hii\ line has meant that it has been 
difficult to study the evolution of $\rm \rhom$ at intermediate redshifts, $z \approx 0.2 - 2.2$. 
Our result, $\rm \rhom = (4.81 \pm 0.75) \times 10^{-4}$, is the first statistically-significant 
estimate of $\rhom$ at these redshifts.

Fig.~\ref{fig:omegahi} shows $\rm \rhom$ versus redshift, from a range of measurements 
at $z \approx 0 - 5$. It is clear that our estimate of $\rm \rhom$ at $z \approx 0.34$ 
is consistent with the value in the local Universe \citep{jones18mnras}, indicating no 
significant evolution in $\rm \rhom$ over $0 < z < 0.4$.

\section{Summary}

We have used a deep uGMRT $970-1370$~MHz observation of the EGS to measure the average 
\hi\ mass and the median total SFR of a sample of 445 blue star-forming galaxies at $z \approx 0.2-0.4$,
by stacking their \hii\ line and 1.4~GHz continuum emission. We obtain $\rm \langle {M}_{HI}\rangle = 
(4.93 \pm 0.70) \times 10^9 \: M_{\odot}$ and $\rm SFR=(0.54 \pm 0.06) \: M_{\odot}\;yr^{-1}$, 
implying an \hi\ depletion timescale of $\rm \langle {\Delta t}_{HI}\rangle \approx \:9\;Gyr$. This is 
comparable to the \hi\ depletion timescale in local star-forming galaxies, but far larger than
the same timescale for star-forming galaxies at $z \approx 1.3$. We obtain a gas fraction of 
$\rm M_{HI}/M_* \approx 1.2$, larger than typical values in local star-forming galaxies. 
Finally, we obtain a cosmic \hi\ mass density of $\rm \rhom = (4.81 \pm 0.75) \times 10^{-4}$, 
consistent with the value of $\rm \rhom$ in the local Universe, and with $\approx 90$\% arising 
from bright galaxies, with $\rm M_B \leq -17$.

\section*{Acknowledgements}

We thank the staff of the GMRT who have made these observations possible. The GMRT is 
run by the National Centre for Radio Astrophysics of the Tata Institute of 
Fundamental Research. NK acknowledges support from the Department of Science and Technology 
via a Swarnajayanti Fellowship (DST/SJF/PSA-01/2012-13).



\end{document}